\documentclass[12pt,aps,pra,groupedaddress,amssymb,amsfonts,nofootinbib,tightenlines]{revtex4} 
\usepackage{graphicx} 

\graphicspath{{./}{graphics/}}
\usepackage{times,color}
\definecolor{purple}{rgb}{0.625,0.125,0.9375}
\newcommand{\ignore}[1]{}

\definecolor{grey}{rgb}{0.8,0.8,0.8}

\ignore{
rm -R /tmp/bench1q; mkdir /tmp/bench1q
cp `texfls bench1q.log | perl -e '$a=<>; $a =~ s:/\S*(revtex4|natbib)\S*(\s|$)::g; print "$a\n"'` /tmp/bench1q
find /tmp/bench1q -type f -name '*.pdf' -exec pdftops -eps {} \; -exec rm {} \;
find /tmp/bench1q -type f -name '*.jpg' -exec perl -e '$f=shift; $g=f$; $g=~s/.jpg/.eps/; exec("convert $f $g");' {} \; -exec rm {} \;
perl -i -n -e 's/^

(cd /tmp/bench1q; tar czvf bench1q.tar.gz *)
}

\ignore{
rm -R /tmp/bench1q; mkdir /tmp/bench1q
cp `texfls bench1q.log | perl -e '$a=<>; $a =~ s:/\S*(revtex4|natbib)\S*(\s|$)::g; print "$a\n"'` /tmp/bench1q
find /tmp/bench1q -type f -name '*.pdf' -exec pdftops -eps {} \; -exec rm {} \;
find /tmp/bench1q -type f -name '*.jpg' -exec perl -e '$f=shift; $g=f$; $g=~s/.jpg/.eps/; exec("convert $f $g");' {} \; -exec rm {} \;
perl -i -n -e 's/^

(cd /tmp/bench1q; tar czvf bench1q.tar.gz *)
}

\newcommand{\ket}[1]{{|}{#1}{\rangle}}

\newcommand{\cG}{{\cal G}}

\newcommand{\cP}{{\cal P}}

\newenvironment{cntrpict}{%
\begin{center}\begin{picture}
}{
\end{picture}\end{center}
}
\newcommand{\nputbox}[3]{\put(#1){\makebox(0,0)[#2]{#3}}}
\newcommand{\nputgr}[4]{\put(#1){\makebox(0,0)[#2]{\includegraphics[#3]{#4}}}}

\color{black}

\begin{document}
\title{Randomized Benchmarking of Quantum Gates}
\author{E. Knill}
\email[]{knill@boulder.nist.gov}
\author{D. Leibfried}
\author{R. Reichle}
\altaffiliation[]{Present address: University of Ulm, Ulm, Germany}
\author{J. Britton}
\author{R. B. Blakestad}
\author{J. D. Jost}
\author{C. Langer}
\altaffiliation[]{Present address: Lockheed Martin, Huntsville, Alabama}
\author{R. Ozeri}
\altaffiliation[]{Present address: Weizmann Institute of Science, Rehovot, Israel}
\author{S. Seidelin}
\author{D. J. Wineland}
\affiliation{National Institute of Standards and Technology} 

\date{\today}
\begin{abstract}
A key requirement for scalable quantum computing is that elementary
quantum gates can be implemented with sufficiently low error.  One
method for determining the error behavior of a gate implementation is
to perform process tomography. However, standard process tomography is
limited by errors in state preparation, measurement and one-qubit
gates. It suffers from inefficient scaling with number of qubits and
does not detect adverse error-compounding when gates are composed in
long sequences. An additional problem is due to the fact that
desirable error probabilities for scalable quantum computing are of
the order of $0.0001$ or lower. Experimentally proving such low errors
is challenging. We describe a randomized benchmarking method that
yields estimates of the computationally relevant errors without
relying on accurate state preparation and measurement. Since it
involves long sequences of randomly chosen gates, it also verifies
that error behavior is stable when used in long computations.  We
implemented randomized benchmarking on trapped atomic ion qubits,
establishing a one-qubit error probability per randomized $\pi/2$
pulse of $0.00482(17)$ in a particular experiment. We expect this
error probability to be readily improved with straightforward
technical modifications.
\end{abstract}

\pacs{03.67.Lx,03.67.Pp,32.80.Pj}

\maketitle

\section{Introduction}

In principle, quantum computing can be used to solve computational
problems having no known efficient classical solutions, such as
factoring and quantum physics simulations, and to significantly speed
up unstructured searches and Monte-Carlo
simulations~\cite{shor:qc1995a,feynman:qc1982a,grover:qc1995a,abrams:qc1999b}.
In order to realize these advantages of quantum computing, we need to
coherently control large numbers of qubits for many computational
steps. The smallest useful instances of the above-mentioned
algorithmic applications require hundreds of qubits and many millions
of steps. A quantum computing technology that realistically can be
used to implement sufficiently large quantum computations is said to
be ``scalable''.  Current quantum computing technologies that promise
to be scalable have demonstrated preparation of nontrivial quantum
states of up to 8 qubits~\cite{haffner:qc2005a}, but it is not yet possible to apply more
than a few sequential two-qubit gates without excessive loss of
coherence. Although there have been experiments to determine the
behavior of isolated gates applied to prepared initial
states~\cite{childs:qc2001b,leibfried:qc2003a,weinstein:qc2004a,obrien:qc2004a,huang:qc2004a,kiesel:qc2005a,haffner:qc2005a,brickman:qc2005a,home:qc2006a,steffen:qc2006a,riebe:qc2007a},
there have been no experiments to determine the noise affecting gates
in a general computational context.

An important challenge of quantum computing experiments is to
physically realize gates that have low error whenever and wherever
they are applied. Studies of fault-tolerant quantum computing suggest
that in order to avoid excessive resource overheads, the probability
of error per unitary gate should be well below $10^{-2}$~\cite{knill:qc2005a,reichardt:qc2004a,raussendorf:qc2006a}. The current
consensus is that it is a good idea to aim for error probabilities
below $10^{-4}$. What experiments can be used to verify such low
error probabilities? One approach is to use process tomography to
establish the complete behavior of a quantum gate. This requires that
the one-qubit gates employed in the tomography have lower error than
the bound to be established on the gate under investigation. If this
requirement is met, process tomography gives much useful information
about the behavior of the gate, but fails to establish that the gate
will work equally well in every context where it may be
required. Process tomography can also be very time consuming as
its complexity scales exponentially with the number of qubits.

We propose a randomized benchmarking method to determine the error
probability per gate in computational contexts.  Randomization has
been suggested as a tool for characterizing features of quantum noise
in~\cite{emerson:qc2005a}. The authors propose implementing random
unitary operators $U$ followed by their inverses $U^{-1}$. Under the
assumption that the noise model can be represented by a quantum
operation acting independently between the implementations of $U$ and
$U^{-1}$, the effect of the randomization is to depolarize the noise.
The average fidelity of the process applied to a pure initial state is
the same as the average over pure states of the fidelity of the noise
operation. (The latter average is known as the average fidelity and is
closely related to the entanglement fidelity of an
operation~\cite{horodecki:qc1999a}.) They also show that the average
fidelity can be obtained with few random experiments.  They then
consider self-inverting sequences of random unitary operations of
arbitrary length. Assuming that the noise can be represented by
quantum operations that do not depend on the choice of unitaries, the
fidelity-decay of the sequence is shown to represent the strength of
the noise.  Our randomized benchmarking procedure simplifies this
procedure by restricting the unitaries to Clifford gates and by not
requiring that the sequence is strictly self-inverting. An alternative
approach to verifying that sequences of gates realize the desired
quantum computation is given in~\cite{magniez:qc2005a}.  In this
approach, successively larger parts of quantum networks are verified
by making measurements involving their action on entangled
states. This ``self testing'' strategy is very powerful and provably
works under minimal assumptions on gate noise. It is theoretically
efficient but requires significantly more resources and multisystem
control than randomized benchmarking.

Our randomized benchmarking method involves applying random sequences
of gates of varying lengths to a standard initial state. Each sequence
ends with a randomized measurement that determines whether the correct
final state was obtained.  The average computationally relevant error
per gate is obtained from the increase in error probability of the
final measurements as a function of sequence length. The random gates
are taken from the Clifford group~\cite{gottesman:qc1998a}, which is
generated by $\pi/2$ rotations of the form $e^{-i\sigma\pi/4}$ with
$\sigma$ a product of Pauli operators acting on different qubits.  The
restriction to the Clifford group ensures that the measurements can be
of one-qubit Pauli operators that yield at least one deterministic
one-bit answer in the absence of errors.  The restriction is justified
by the fact that typical fault-tolerant architectures (those based on
stabilizer codes) are most sensitive to errors in elementary Clifford
gates such as the controlled NOT. Provided the errors in these gates
are tolerated, other gates needed for universality are readily
implemented~\cite{bravyi:qc2005a,knill:qc2005a}.  Note that the
results of~\cite{emerson:qc2005a} hold if the unitaries are restricted
to the Clifford group, because the Clifford group already has the
property that noise is depolarized. We believe that randomized
benchmarking yields computationally relevant errors even when the
noise is induced by, and depends on, the gates, as is the case in
practice.

Randomized benchmarking as discussed and implemented here gives an
overall average fidelity for the noise in gates.  To obtain more
specific information, the technique needs to be refined.
In~\cite{levi:qc2006a}, randomization by error-free one-qubit
unitaries is used to obtain more detailed information about noise
acting on a multiqubit system. Randomized benchmarking can be adapted
to use similar strategies.

\section{Randomized benchmark of one qubit}

For one qubit, our randomized benchmarking procedure consists of a
large number of experiments, where each experiment consists of a pulse
sequence that requires preparing an initial quantum state $\rho$,
applying an alternating sequence of either major-axis $\pi$ pulses or
identity operators (``Pauli randomization'') and $\pi/2$ pulses
(``computational gates''), and performing a final measurement $M$.
The pulse sequence between state preparation and measurement begins
and ends with $\pi$ pulses.  For one qubit, the initial state is
$\ket{0}$. Because the major-axis $\pi$- and $\pi/2$ rotations are in
the Clifford group, the state is always an eigenstate of a Pauli
operator during the pulse sequence. The Pauli randomization applies
unitary operators (``Pauli pulses'') that are (ideally) of the form
$e^{\pm i\sigma_b\pi/2}$, where the sign $\pm$ and $b=0,x,y,z$ are
chosen uniformly at random and we define $\sigma_0$ to be the identity
operator.  For ideal pulses, the choice of sign determines only a
global phase. However, in an implementation, the choice of sign can
determine a physical setting that may affect the error behavior.  The
computational gates are $\pi/2$ pulses of the form $e^{\pm
i\sigma_u\pi/4}$, with $u=x,y$. The sign and $u$ are chosen uniformly
at random, except for the last $\pi/2$ pulse, where $u$ is chosen so
that the final state is an eigenstate of $\sigma_z$.  The
computational gates generate the Clifford group for one qubit. Their
choice is motivated by the fact that they are experimentally
implementable as simple pulses.  The final measurement is a
von Neumann measurement of $\sigma_z$.  The last $\pi/2$ pulse ensures
that, in the absence of errors, the measurement has a known,
deterministic outcome for a given pulse sequence. However, the
randomization of the pulse sequence ensures that the outcome is not
correlated with any individual pulse or proper subsequence of pulses.

The length $l$ of a randomized pulse sequence is its number of $\pi/2$
pulses. The $\pi/2$ pulses are considered to be the ones that advance
a computation. The $\pi$ pulses serve only to randomize the
errors. One can view their effect as being no more than a change of
the Pauli frame. The Pauli frame consists of the Pauli operator that
needs to be applied to obtain the intended computational state in the
standard basis~\cite{knill:qc2005a}. We call the $\pi/2$ and Pauli
pulse combinations randomized computational gates.  In principle, we
can determine a pulse error rate by performing $N$ experiments for
each length $l=1,\ldots,L$ to estimate the average probability $p_l$
of the incorrect measurement outcome (or ``error probability'') for
sequences of length $l$. The relationship between $l$ and $p_l$ can be
used to obtain an average probability of error per pulse.  Suppose
that all errors are independent and depolarizing.  Let the
depolarization probability of an operation $A$ be $d_A$ and consider a
specific pulse sequence consisting of operations
$A_0,A_1A_2,\ldots,A_{2l+1}A_{2l+2},A_{2l+3}$, where $A_0$ is the
state preparation, $A_1A_2$ and the following pairs are the randomized
computational gates, and $A_{2l+3}$ the measurement.  For the
measurement, we can assume that the error immediately precedes a
perfect measurement.  The state after $A_k$ is a known eigenstate of a
Pauli operator or completely depolarized. Depolarization of the state
is equivalent to applying a random Pauli or identity operator, each
with probability $1/4$.  The probability of the state's not having
been depolarized is $\prod_{j=0}^k(1-d_{A_j})$. In particular, we can
express $p_l = E((1-\prod_{j=0}^{2l+3} (1-d_{A_j}))/2)$, where the
function $E(.)$ gives the expectation over the random choices of the
$A_j$. The factor of $1/2$ in the expression for $p_l$ arises because
depolarization results in the correct state $1/2$ of the time.  The
choices of the $A_j$ are independent except for the last $\pi/2$
pulse.  Assume that the depolarization probability of the last $\pi/2$
pulse does not depend on the previous pulses. We can then write $p_l =
(1-(1-d_{\mathrm{if}})(1-d)^l)/2$, where $d$ is the average
depolarization probability of a random combination of one $\pi/2$- and
one Pauli pulse (a randomized computational gate), and
$d_{\mathrm{if}}$ combines the depolarization probabilities of the
preparation, initial Pauli pulse and measurement. Thus $p_l$ decays
exponentially to $1/2$, and the decay constant yields $d$.

A commonly used metric to describe the deviation of an
implemented gate from the intended gate is the average fidelity $F_a$,
which is defined as the uniform average over pure input states of the
fidelity of the output state with respect to the intended output state.  We are
interested in the average computationally relevant error per step
consisting of a randomized computational gate (``average error'' for
short). This is given by the expectation over gates of $1-F_a$ and
relates to the depolarization parameter $d$ of the previous paragraph
by $1-F_a=d/2$.  In our implementation of the randomized computational
gates, the $\pi$ pulses around the $z$-axis are implemented by changes
in rotating frame and do not involve actively applying a pulse.
Therefore, on average, the angular distance of the randomized gate's
action is $\pi$. As a result, $(1-d/2)$ represents the average
fidelity of pulses with action $\pi$.

Although estimates of $p_l$ are sufficient to obtain the average error
for a randomized computational gate, it is useful to
consider the error behavior of specific randomized computations and
even fixed instances of the randomized sequences.  For this purpose,
the sequences are generated by first producing $N_G$ random sequences
consisting of $L$ random computational gates, where the gates are
chosen independently without considering the final state. These
sequences are considered to be a sample of typical computations. Each
sequence is then truncated at different lengths. For each length, a
$\pi/2$ pulse is appended to ensure that the final state is an
eigenstate of $\sigma_z$. The sign of this final pulse is
random.  The resulting sequences are randomized by
inserting the random Pauli pulses. We can then perform experiments to
determine the probability of incorrect measurement outcomes for each
such sequence and for each truncated computation after randomization
by Pauli pulses. To be specific, the procedure is implemented as
follows:
\begin{list}{}{}
\item[\textbf{Randomized benchmarking for one qubit:}] This obtains
measurement statistics for $N_G  N_l N_P N_e$ experiments, where $N_G$ is
the number of different computational gate sequences, $N_l$ is the
number of lengths to which the sequences are truncated, $N_P$ is the
number of Pauli randomizations for each gate sequence, and $N_e$ is
the number of experiments for each specific sequence. 
\item[1.] Pick a set of lengths $l_1<l_2<\ldots<l_{N_l}$. The goal
is to determine the probability of error of randomized 
computations of each length.
\item[2.] Do the following for each $j=1,\ldots,N_G$:
\begin{list}{}{}
  \item[2.a.] Choose a random sequence $\cG=\{G_1,\ldots\}$ of $l_{N_l}-1$ 
  computational gates.
  \item[2.b.] For each $k=1,\ldots,N_l$ do the following:
  \begin{list}{}{}
    \item[2.b.1.] Determine the final state $\rho_f$ obtained by
    applying $G_{l_k}\ldots G_1$ to $\ket{0}$, assuming no error.
    \item[2.b.2.] Randomly pick a final computational gate $R$
    among the two $\pm x, \pm y, \pm z$ axis $\pi/2$ pulses that result in an eigenstate
    of $\sigma_z$ when applied to $\rho_f$. Record which eigenstate is
    obtained.
    \item[2.b.3.] Do the following for each $m=1,\ldots N_P$:
    \begin{list}{}{}
      \item[2.b.3.a] Choose a random sequence $\cP=\{P_1,\ldots\}$ 
      of $l_k+2$ Pauli pulses.
      \item[2.b.3.b.] Experimentally implement the pulse sequence
      that applies $P_{l_k+2}RP_{l_k+1}G_{l_k}\ldots G_1P_1$ to $\ket{0}$
      and measures $\sigma_z$, repeating the experiment $N_e$ times.
      \item[2.b.3.c.] From the experimental data and the expected
      outcome of the experiments in the absence of errors (from step 2.b.2), 
      obtain an estimate $p_{j,l_k,m}$ of the probability of error.
      Record the uncertainty of this estimate.
    \end{list}
  \end{list}
\end{list}
\end{list}
The probabilities of error $p_l$ are obtained from the $p_{j,l_k,m}$
by averaging $p_{l_k} =
\sum_{j=1}^{N_G}\sum_{m=1}^{N_P}p_{j,l_k,m}/(N_G N_P)$.  We also obtain
the probabilities of error for each computational gate sequence,
$p_{j,l_k} = \sum_{m=1}^{N_P}p_{j,l_k,m}/N_P$. If the errors are
independent and depolarizing, the $p_{j,l_k,m}$ and the $p_{j,l_k}$
should not differ significantly from the $p_{l_k}$. However, if the
errors are systematic in the sense that each implemented pulse differs
from the ideal pulse by a pulse-dependent unitary operator, this can
be observed in the distribution of the $p_{j,l_k,m}$ over $m$.  In
this case, the final state of each implemented pulse sequence is pure.
The deviation of these pure states from the expected states is
distributed over the Bloch sphere as $m$ and $j$ are varied. For
example, consider the case where $p_{l_k}$ is close to $1/2$. If the errors
are systematic, the $p_{j,l_k,m}$ are distributed as the probability amplitude of
$\ket{1}$ for a random pure state. In particular, we are likely to
find many instances of $j$ and $m$ where $p_{j,l_k,m}$ is close to $0$
or $1$, that is, differs significantly from $1/2$. In contrast, if the
error is depolarizing, the $p_{j,l_k,m}$ are all close to $1/2$
independent of $j$ and $m$.

\section{Trapped-ion-qubit implementation}

We determined the computationally relevant error probabilities for
computational gates on one qubit in an ion trap.  The qubit was
represented by two ground-state hyperfine levels of a
${}^9\textrm{Be}^+$ ion trapped in a linear radio-frequency Paul trap
briefly described in~\cite{wineland:qc2005a}. It is the same trap that
has been used in a several quantum information processing
experiments~\cite{barrett:qc2004a,chiaverini:qc2004a,chiaverini:qc2005a,leibfried:qc2005a,reichle:qc2006a}.
The two qubit states are $\ket{\downarrow}$ ($F=2,m_F=-2$) and
$\ket{\uparrow}$ ($F=1,m_F=-1$), where for our purposes, we identify
$\ket{\downarrow}$ with $\ket{0}$ and $\ket{\uparrow}$ with $\ket{1}$.
The state $\ket{\downarrow}$ is prepared by optical pumping, after
laser cooling the motional states of the ion.  We can distinguish
between $\ket{\downarrow}$ and $\ket{\uparrow}$ by means of
state-dependent laser fluorescence. Computational gates and Pauli
pulses involving $x$- or $y$-axis rotations were implemented by means
of two-photon stimulated Raman transitions. To ensure that the pulses
were not sensitive to the remaining excitations of the motional
degrees of freedom, we used copropagating Raman beams. It was
therefore not necessary to cool to the motional ground state and only
Doppler cooling was used. Pulses involving $z$-axis rotations were
implemented by programmed phase changes of one of the Raman
beams. This changes the phase of the rotating reference frame and is
equivalent to the the desired $z$-axis rotation.  The $z$-axis
rotations were accompanied by a delay equivalent to the correponding
$x$ and $y$ pulses.

The Raman beams were switched on and off and shifted in phase and
frequency as necessary by means of acousto-optic modulators
controlled by a field-programmable gate array (FPGA). The pulse sequences
were written in a special-purpose pulse-programming language and
precompiled onto the FPGA. The version of the FPGA in use for the
experiments was limited to about 100 computational pulses. The longest
sequence in our experiments consisted of 96 computational gates. Our
initial implementations clearly showed the effects of systematic
errors in the distribution of the error probabilities of individual
sequences.  This proved to be a useful diagnostic and we were able to
correct these systematics to some extent. One of the largest
contributions to systematic errors was due to Stark shifts.  To
correct for for these shifts, we calibrated them and adjusted phases
in the pulse sequences.

\section{Experimental results}

\begin{figure}
\begin{cntrpict}(3,3)(0,-3)
\nputgr{0,0}{tl}{width=3in}{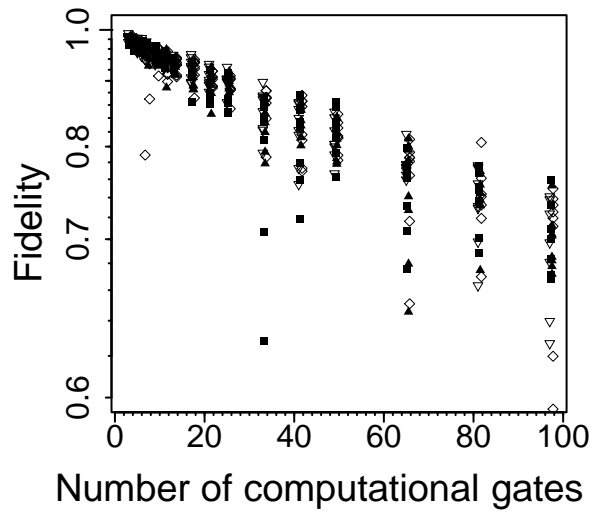}
\nputbox{-.1,-1.5}{l}{\rotatebox{90}{\large\textsf{Fidelity}}}
\nputbox{1.5,-3}{b}{\large\textsf{Number of computational gates}}
\end{cntrpict}
\caption{Fidelity as a function of the number of steps for each
randomized sequence. The fidelity ($1-\textrm{prob.~of error}$) is
plotted on a logarithmic scale. The fidelity for the final state is
measured for each randomized sequence. There are $32$ points for each
number of steps, corresponding to $8$ randomizations of each of four
different computational sequences. Different symbols are used for the
data for each computational sequence.  The standard error of each
point is between $0.001$ (near fidelities of $1$) and $0.006$ (for the
smaller fidelities). The scatter greatly exceeds the standard error,
suggesting that coherent errors contribute significantly to the loss
of fidelity.}
\label{fig:each_all}
\end{figure}

\begin{figure}
\begin{cntrpict}(3,3)(0,-3)
\nputgr{0,0}{tl}{width=3in}{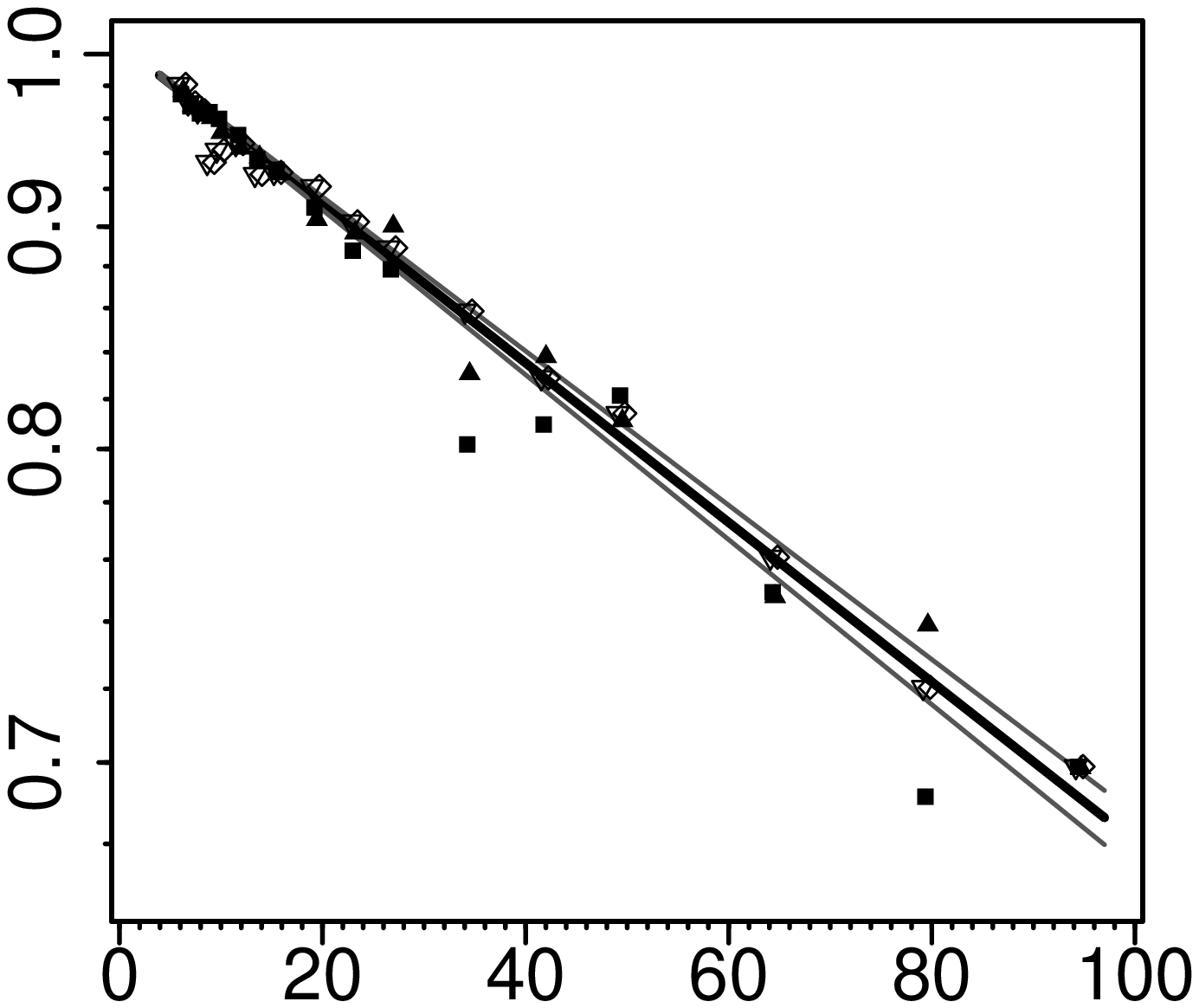}
\nputbox{-.15,-1.5}{l}{\rotatebox{90}{\large\textsf{Average fidelity}}}
\nputbox{1.5,-3}{b}{\large\textsf{Number of computational gates}}
\end{cntrpict}
\caption{Average fidelity as a function of the number of steps for
each computational sequence.  The points show the average randomized
fidelity for four different computational gate sequences (indicated by
the different symbols) as a function of the length.  The average fidelity is
plotted on a logarithmic scale. The middle line shows the fitted exponential
decay. The upper and lower line show the boundaries of the $68\,\%$
confidence interval for the fit. The standard deviation of each point
due to measurement noise ranges from $0.0004$
for values near $1$ to $0.002$ for the lower values, smaller than the
size of the symbols. The empirical standard deviation based on the
scatter in the points shown in Fig.~\ref{fig:each_all} ranges from
$0.0011$ to $0.014$.
The slope implies an error probability of
$0.00482(17)$ per randomized computational gate. The data is consistent
with the gate's errors not depending on position in the sequence.}
\label{fig:merged_allwl}
\end{figure}

We generated $N_G=4$ random computational sequences and truncated them
to the $N_l=17$ lengths
$\{2,3,4,5,6,8,10,12,16,20,24,32,40,48,64,80,96\}$.  Each truncated
sequence was Pauli randomized $N_P=8$ times. Each final pulse sequence
was applied to an ion a total of $8160$ times in four groups that
were interleaved with the other experiments in a randomized
order. Pulse durations, qubit-resonant frequencies and Stark shifts
were recalibrated automatically at regular intervals. The number of
experiments per pulse sequence was sufficient to obtain the
probability of incorrect measurement outcome with a statistical error
small compared to the variation due to randomization and systematic
errors. Fig.~\ref{fig:each_all} plots the fidelity (one minus the
probability of incorrect measurement outcome) of each of the
$4*17*8=544$ final pulse sequences against the length of the
corresponding computational sequence. As explained in the figure caption,
the variation in fidelity for
each length shows that non-depolarizing errors contribute
significantly to error. Fig.~\ref{fig:merged_allwl} plots the average
fidelity over the eight Pauli randomizations of each computational
sequence truncated to the different lengths.  Pauli randomization
removes coherent errors, significantly reducing the variation in
fidelities for different computational sequences.  The remaining
variation could be due to the small sample of $8$ Pauli randomizations
used to obtain the average. The empirical average probability of error
per randomized computational gate can be obtained by fitting the
exponential decay and was found to be $0.00482(17)$.  The fit was
consistent with a simple exponential decay, which suggests that these
gates behave similarly in all computational contexts.  The error bars
represent standard deviation as determined by
nonparameteric bootstrapping~\cite{efron:qc1993a}. In what follows,
if the fits are good, error bars are determined
from nonlinear least-squares fits. In the cases where we can obtain a
useful estimate of an error per randomized computational gate but the
fits are poor, we used nonparameteric bootstrapping.

For our experimental setting, it is possible to perform experiments to
quantify the different types of errors as a consistency check. The
results of these experiments are in App.~\ref{app:checks}
and are consistent with the randomized benchmarking data.

\section{Theoretical considerations}

The average error per randomized computational gate is obtained by
fitting an exponential. For general error models, it is possible that
the initial behavior of the measured error probabilities does not
represent the average error of interest, and it is the eventual decay
behavior that is of interest. In this case, randomized benchmarking
determines an asymptotic average error probability (AAEP) per randomized
computational gate. It is desirable to relate the empirical
AAEP to the average error probability (AEP) of a single randomized
computational gate. As discussed above, the AAEP agrees with the
AEP if the error of all operations is depolarizing and independent of
the gates. It can be seen that for depolarizing errors, this
relationship holds even if the error depends on the gates.  In
general, one can consider error models with the following properties:
\begin{list}{}{}
\item[]\textbf{Memoryless errors.} The errors of each gate
are described by a quantum operation. In particular, the ``environment''
for errors in one gate is independent of that in another.
\item[]\textbf{Independent errors.} For gates acting in parallel
on disjoint qubits, each gate's errors are described by a quantum operation
acting on only that gate's qubits.
\item[]\textbf{Stationary errors.} The errors depend only on the
gate, not on where and when in the process
the error occurs.
\item[]\textbf{Subsystem preserving errors.}
The errors cause no leakage out of the subsystem defining the
qubits.
\end{list}
Although the AAEP need not be identical to the AEP, we conjecture that
there are useful bounds relating the two error probabilities. In
particular, if the AAEP is zero then there is a fixed logical frame in
which the AEP is zero. Trivially, if the AEP is zero, then the AAEP is
zero.

Randomized benchmarking involves both Pauli randomization and
computational gate randomization. The expected effect of Pauli
randomization is to ensure that, to first order, errors consist of
random (but not necessarily uniformly random) Pauli operators.
Computational gate randomization ensures that we average errors over
the Clifford group. If, as in our experimental implemetation, the
computational gates generate only the Clifford group, it takes a few
steps for the effect to be close to averaging over the Clifford
group. This process is expected to have the effect of making all
errors equally visible to our measurement, even though the measurement
is fixed in the logical basis and the last step of the randomized
computation is picked so that the answer is deterministic in the
absence of errors.

\section{Benchmarking mutliple qubits}

Scalable quantum computing requires not only having access to many
qubits, but also the ability to apply many low-error quantum gates to
these qubits. The error behavior of gates should not become worse as
the computation proceeds. Randomized benchmarking can verify the
ability to apply many multiqubit gates consistently.

Randomized benchmarking can be applied to two or more qubits by
expanding the set of computational gates to include multiqubit gates.
The initial state is $\ket{0\ldots 0}$.
Pauli randomization is performed as before and is expected to convert
the error model to probabilistic Pauli errors to first order.  Because
the size of the Clifford group for two or more qubits is large, one
cannot expect to effect a random Clifford group element at each
step. Instead, one has to rely on rapid mixing of random products of
generators of the Clifford group to achieve (approximate) multiqubit
depolarization.  The number of computational steps that is required
for approximate depolarization depends on the computational gate
set. An example of a useful gate set consists of controlled NOTs
(alternatively, controlled sign flips) combined with major-axis
$\pi/2$ pulses on individual qubits.  By including sufficiently many
one-qubit variants of each gate, one can ensure that each step's
computational gates are randomized in the product of the one-qubit
Clifford groups.  This already helps: It has the effect of equalizing
the probability of Pauli product errors of the same weight
(see~\cite{levi:qc2006a}).

The one-qubit randomized benchmark has a last step that ensures a
deterministic answer for the measurement. For $n>1$ qubits, one cannot
expect deterministic answers for each qubit's measurement,
as this may require too complex a
Clifford transformation. Instead, one can choose a random Pauli
product that stabilizes the last state and apply a random product of
one-qubit $\pi/2$ pulses with the property that this Pauli product is
turned into a product of $\sigma_z$ operators.  If there is no error,
measuring $\sigma_z$ for each qubit and then computing the appropriate
parity of the measurement outcomes gives a known deterministic
answer. With error, the probability of obtaining the wrong parity can
be thought of as a one-qubit error probability $p$ for the
sequence. If the error is completely depolarizing on all qubits, with
depolarization probability $d$, then $p = d/2$, just as for one
qubit. One expects that for sufficiently long sequences, $p$ increases
exponentially toward $1/2$ so that the asymptotic average error
probability per randomized computational gate can be extracted as for
one qubit.

\begin{acknowledgments}
This work was supported by DTO and NIST.  It is a contribution of the
National Institute of Standards and Technology, an agency of the
U.S. government, and is not subject to U.S. copyright.
\end{acknowledgments}

\appendix

\section{Direct Error Characterizations}
\label{app:checks}

We performed experiments to directly quantify the different types of
errors in our pulses. These experiments characterize only the initial
error (the error of the first gates) and serve as a consistency check
for the randomized benchmarking data.

Known sources of errors include (a) phase errors due to fluctuating
magnetic fields and changes in path length between the two Raman beams
(they are merged on a polarizing beamsplitter before targetting the
ion), (b) amplitude errors due to changes in beam position at the ion and
intensity fluctuations not compensated by the ``noise eaters'' (active
beam intensity stabilization), and (c) spontaneous emission from the upper
levels required for the stimulated Raman transition.

\begin{figure}
\begin{cntrpict}(3,3)(0,-3)
\nputgr{0,0}{tl}{width=3in}{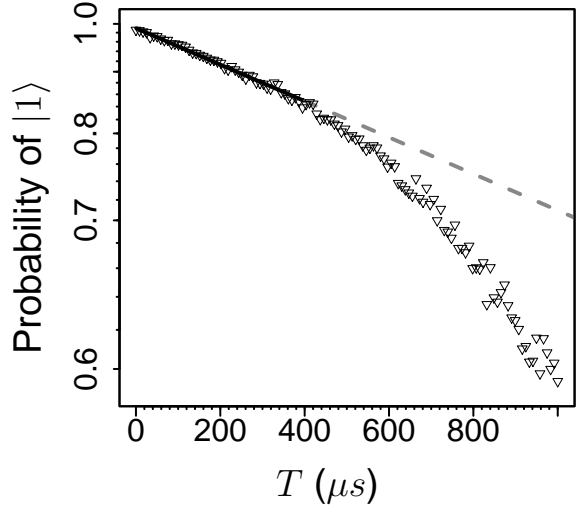}
\nputbox{-.15,-1.5}{l}{\rotatebox{90}{\large\textsf{Probability of $\ket{1}$}}}
\nputbox{1.5,-3}{b}{\large\textsf{$T$ ($\mu s$)}}
\end{cntrpict}
\caption{Measurement of phase decoherence with refocusing.  We
measured the probability of $\ket{1}$ as a function of time for the
standard refocused decoherence measurement.  The pulse sequence
consisted of a $\pi/2$ pulse at phase $0$ followed by a delay of
$T/2$, a $\pi$ pulse at phase $\pi$, another delay of $T/2$ and a
final $\pi/2$ pulse at phase $\pi$.  The straight line shows the fit
for exponential decay on the interval from $1$ to $200\,\mu s$.  Its
extrapolation to larger times is shown dashed. The deviation from an
exponential decay at larger times can be attributed to slow phase
drifts that are no longer refocused by the single $\pi$ pulse in
the pulse sequence.  From the fit, the contribution of unrefocusable
phase decoherence to the error probability per step is $0.0037(1)$.
The standard deviation of the plotted points ranges from $0.002$ for
values near $1$ to $0.008$ for the smallest values, similar to the
apparent scatter of the plotted points.
}
\label{fig:cocar_decoh}
\end{figure}

\begin{figure}
\begin{cntrpict}(3,3)(0,-3)
\nputgr{0,0}{tl}{width=3in}{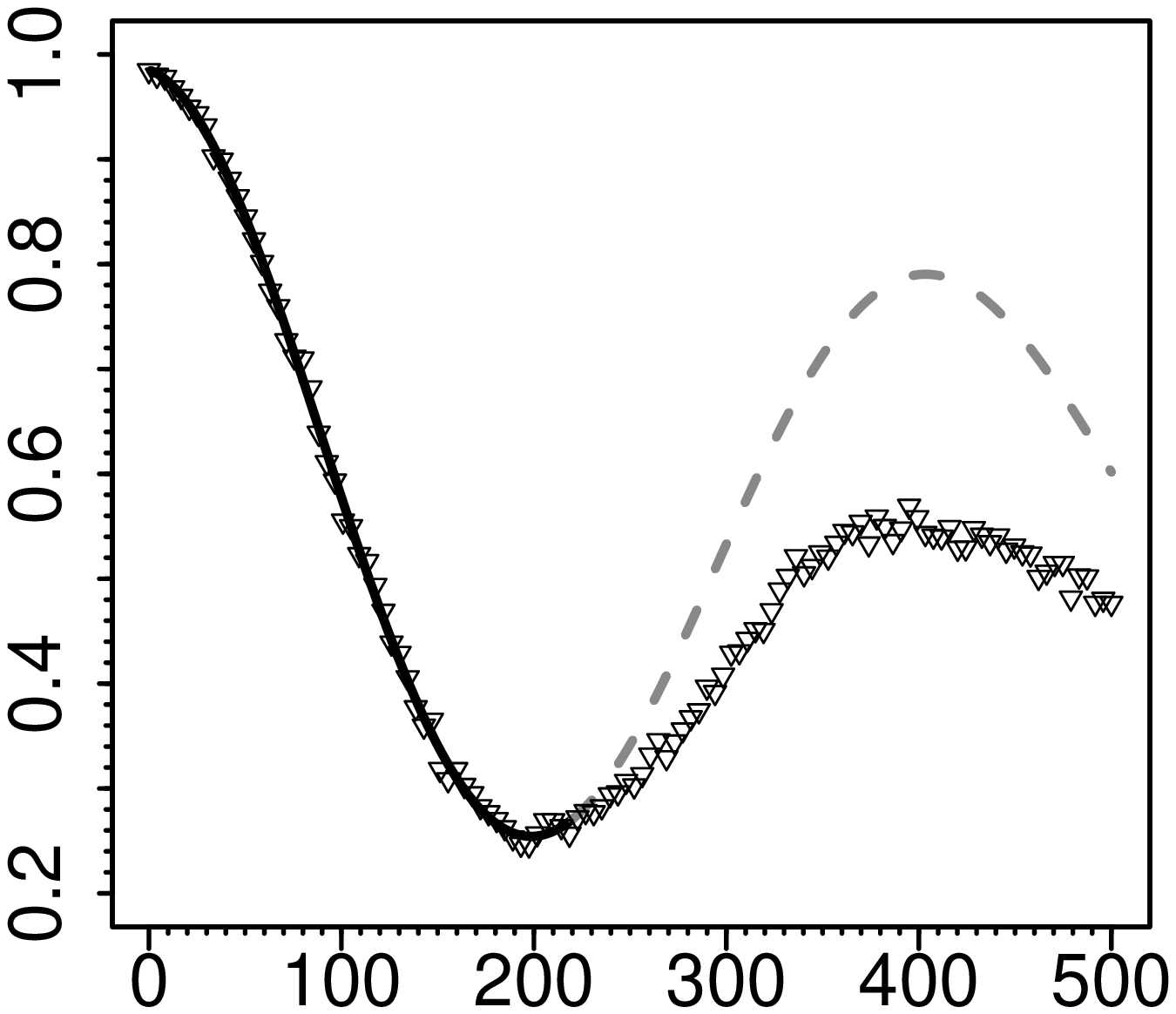}
\nputbox{-.1,-1.5}{l}{\rotatebox{90}{\large\textsf{Probability of $\ket{1}$}}}
\nputbox{1.5,-3}{b}{\large\textsf{$\mu$s}}
\end{cntrpict}
\caption{Measurement of phase decoherence without refocusing.  The
randomized benchmark does not systematically refocus changes in
frequency. To estimate the contribution to error from decoherence
including refocusable decoherence, we performed the experiment of
Fig.~\ref{fig:cocar_decoh} without the refocusing pulse.  This is
essentially an on-resonance Ramsey experiment. It was not
experimentally possible to eliminate the oscillatory shape of the
curve by calibrating the frequency indicating that the oscillation was
not simply caused by detuning from the resonant frequency, However,
the shape is similar to what one would expect from a roughly periodic
change in frequency that is not synchronized with the experiment. Such
changes could come from magnetic field fluctuations and phase noise
due to air currents in the paths of the two Raman beams. To estimate
the contribution to the probability of error per step, we fitted an
exponentially decaying $\cos(t)$ curve to the points with time
coordinates less than $220$ $\mu$s. The extrapolation of the fitted
curve (dashed) clearly deviates from the data. Note that for
sinusoidal phase noise, the curve should be related to a decaying
Bessel function. Fits to such a function also deviate from the
experimental data, consistent with the phase fluctuations not being
sinusoidal.  Since the contribution to the probability of error is
derived from the short-time behavior, the effect of the different
models on the inferred probability of error per step is small.
For the fit shown, the inferred contribution to the probability of
error per step is $0.0090(7)$, larger than the error per step derived
from Fig.~\ref{fig:merged_allwl}. This is likely due to the fact that
in the randomized sequences, the centering of the explicit $\pi$
pulses in their intervals reduces this contribution by refocusing.  }
\label{fig:cocar_unref}
\end{figure}

Phase decoherence can be measured by observing the decay of signal in
a Ramsey spectrometry experiment of the qubit with or without
refocusing~\cite{ozeri:qc2005a}.  Fig.~\ref{fig:cocar_decoh} shows the
probability of observing $\ket{1}$ at the end of a refocused Ramsey
experiment as a function of the delay between the first and last
$\pi/2$ pulse.  By fitting the initial part of the curve to an
exponential decay, one can infer the contribution of unrefocusable
phase error to each step of the Pauli randomized sequences. We
obtained an estimate of $0.0037(1)$ for this contribution.
Fig.~\ref{fig:cocar_unref} shows the probability of observing
$\ket{1}$ in a similar experiment but with the refocusing pulse
omitted. This is an on-resonance Ramsey experiment.  The fit suggests
a contribution of $0.0090(7)$ for the error per step.  This is larger
than the inferred error from the randomized experiments, which can be
explained by the refocusing effects of the Pauli randomization.  See
the caption of Fig.~\ref{fig:cocar_unref} for a discussion of fitting
issues.  We note that our benchmarking experiments, as well as the
error characterizations in this section, were performed without
line-triggering the experiments, thereby making them sensitive to
phase shifts caused by $60\,\textrm{hz}$ magnetic field
fluctuations. Greatly improved decoherence times are typically
obtained if such triggering is used.

\begin{figure}
\begin{cntrpict}(3,3)(0,-3)
\nputgr{0,0}{tl}{width=3in}{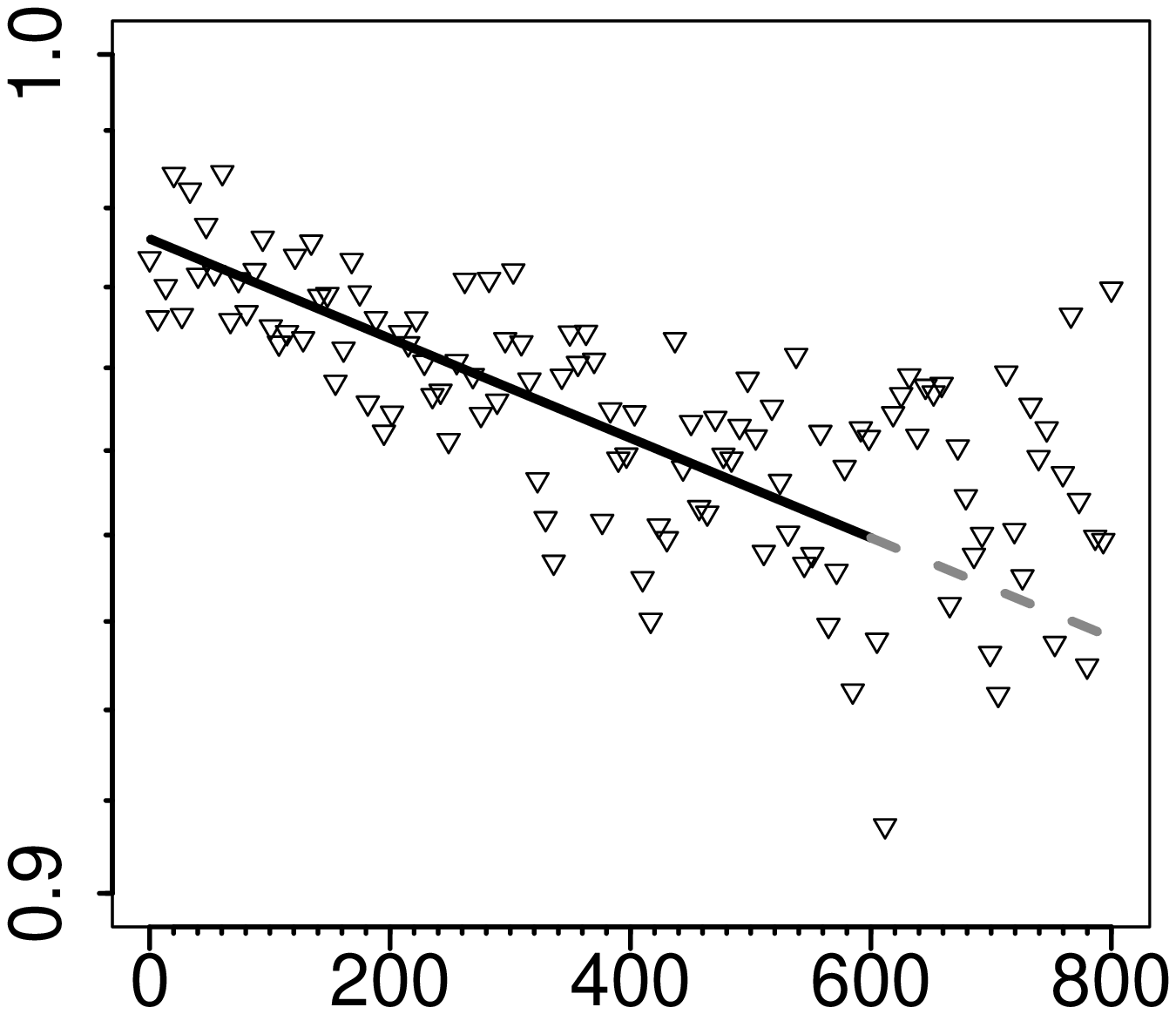}
\nputbox{-.1,-1.5}{l}{\rotatebox{90}{\large\textsf{Probability of $\ket{1}$}}}
\nputbox{1.5,-3}{b}{\large\textsf{$\mu$s}}
\end{cntrpict}
\caption{Contribution of spontaneous emission to phase decoherence.
To experimentally determine the contribution of spontaneous emission
to decoherence, we applied the two Raman beams separately for half the
time of each arm of the refocused decoherence measurement and compared
the resulting data to that of
Fig.~\ref{fig:cocar_decoh}~\cite{ozeri:qc2005a}.  The points shown
here were obtained by dividing the probabilities measured by the
corresponding probabilities of Fig.~\ref{fig:cocar_decoh},
interpolating between the nearest points to match the time
coordinates. The straight line shows the fitted exponential decay. The
fit was weighted and used linear approximation to determine standard
deviations of the points.  The standard deviations used range from
$0.003$ to $0.015$, which is substantially less than the apparent
scatter of the plotted points.
The inferred contribution to the error probability per step is $0.00038(3)$.
This contribution can be estimated theoretically~\cite{ozeri:qc2005a}, which 
for the relevant configuration gives a value of approximately $0.0003$.
}
\label{fig:cocar_spont}
\end{figure}

The contribution of spontaneous emission to phase decoherence can be
determined by a refocused Ramsey experiment where the two Raman beams
are on separately half the time during the intervals between the
pulses~\cite{ozeri:qc2005a}. To determine the desired contribution, the probability of
$\ket{1}$ as a function of time is compared to the data shown in
Fig.~\ref{fig:cocar_decoh}. The results of the comparisons are
in Fig.~\ref{fig:cocar_spont}. The inferred contribution to the error
probability per step is $0.00038(3)$, well below the contribution
of the other sources of error.

\begin{figure}
\begin{cntrpict}(3,3)(0,-3)
\nputgr{0,0}{tl}{width=3in}{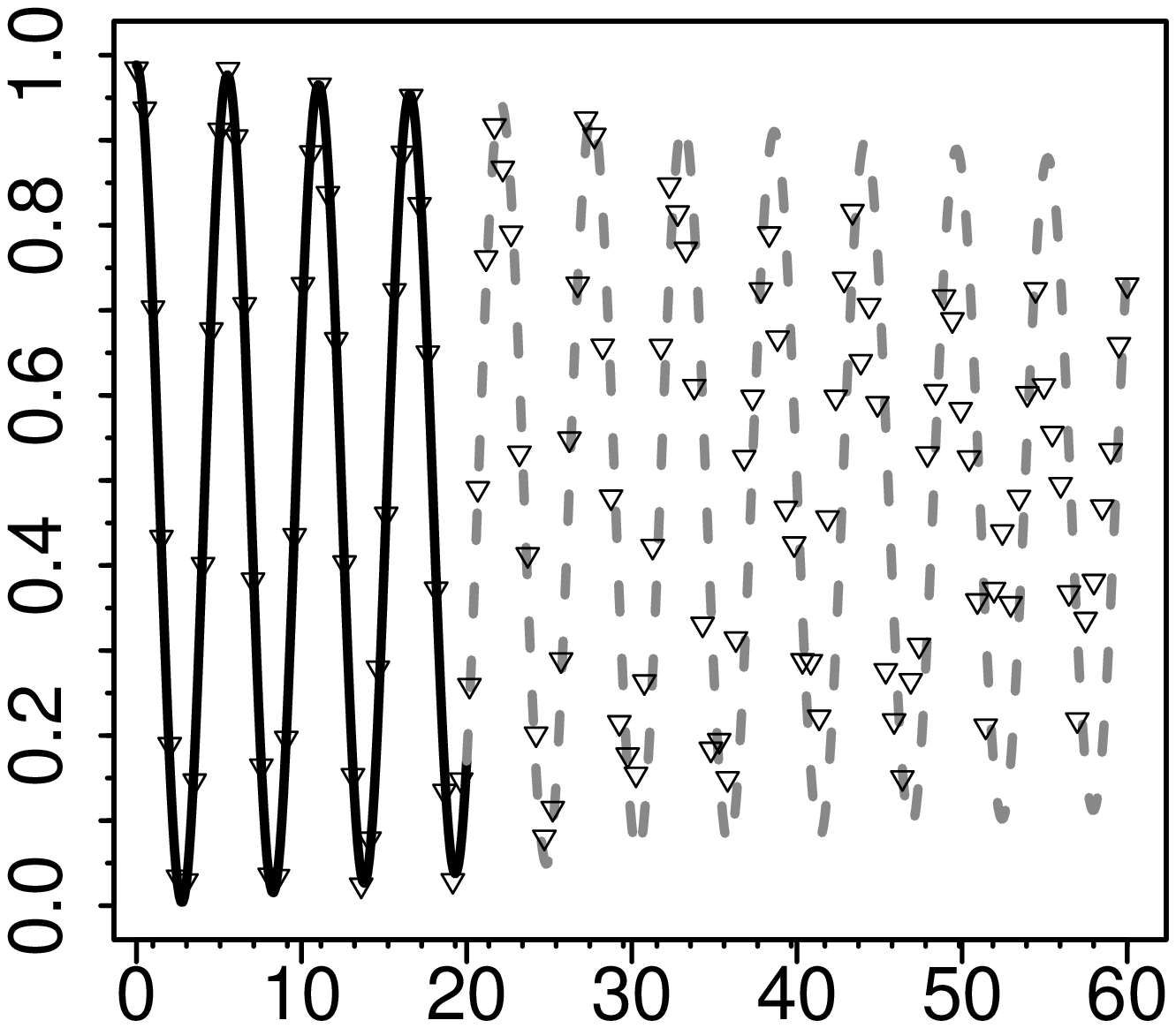}
\nputbox{-.1,-1.5}{l}{\rotatebox{90}{\large\textsf{Probability of $\ket{1}$}}}
\nputbox{1.5,-3}{b}{\large\textsf{$\mu$s}}
\end{cntrpict}
\caption{Rabi flopping experiment.  To determine the contribution to
the probability of error per step due to pulse area error and associated
decoherence, we performed a Rabi flopping experiment. We fitted the
points to a decaying cosine curve with a possible phase offset and
both linear and quadratic decay.  Again, we restricted the fit to an
initial segment of the data (black curve). The extrapolation (dashed
curve) shows significant deviations. The random uncertainty in the points
ranges from $0.002$ to $0.007$, less than the symbol size of the plotted
points. 
The apparent scatter in the points near the end of the curve
is likely due to slow fluctuations in pulse amplitude.
The contribution to the probability of error per step as 
detected in this experiment is $0.006(3)$ if the calibration were
based on this experiment. Automatically calibrated
pulse times fluctuated by around $0.02$ $\mu$s.
For pulse times differing by this amount,
the contribution to the error per step is $0.007(3)$.
}
\label{fig:cocar_rabi}
\end{figure}

The effect of amplitude fluctuations can be estimated from the loss of
visibility of a Rabi flopping experiment.  The data are shown in
Fig.~\ref{fig:cocar_rabi}. Modeling the Rabi flopping curve is
non-trivial and the fits are not very good. Nevertheless, we can
estimate a contribution to the error probability per step from the
behavior of the curve during the first few oscillations.  This gives a
contribution of $0.006(3)$, consistent with the probability of error
per step obtained in the randomized experiments. Note that
the contribution measured here also includes errors due to phase
fluctuations during the computation pulses.

\bibliography{journalDefs,bench1q}

\end{document}